\documentclass[aps,epsf,subfigure,twocolumn,prl]{revtex4-1}

\usepackage{graphicx}
\usepackage{epstopdf}
\usepackage{dcolumn}
\usepackage{bm}
\usepackage{amsmath}
\usepackage{color,psfrag,epsfig}

\input{Qcircuit}

\begin{document}

\title{Quantum Error Correction with magnetic molecules}
\author{Jos\'e J. Baldov\'\i,$^1$ Salvador Cardona-Serra,$^1$ Juan M.
Clemente-Juan,$^1$ Luis Escalera-Moreno,$^1$ Alejandro Gaita-Ari\~no,$^{1*}$
Guillermo M\'\i{}nguez Espallargas$^1$}
\affiliation{$^1$Instituto de Ciencia Molecular (ICMol), Universidad de Valencia,
Catedr\'atico Jos\'e Beltr\'an 2, 46980 Paterna, Spain}
\date{\today}

\begin{abstract}
Quantum algorithms often assume independent spin qubits to produce trivial
$|\uparrow\rangle=|0\rangle$, $|\downarrow\rangle=|1\rangle$ mappings. This can
be unrealistic in many solid-state implementations with sizeable magnetic
interactions. Here we show that the lower part of the spectrum of a
molecule containing three exchange-coupled metal ions with $S=1/2$ and $I=3/2$
is equivalent to nine electron-nuclear qubits. We derive the relation between
spin states and qubit states in reasonable parameter ranges for the rare earth
Tb$^{3+}$ and for the transition metal Cu$^{2+}$ , and study the possibility to
implement Shor's Quantum Error Correction code on such a molecule. We also
discuss recently developed molecular systems that could be adequate from an
experimental point of view.
\end{abstract}

\maketitle

{\it Introduction} --
A key problem of quantum computing is derived from the no-cloning
theorem~\cite{noclon} which states that an unknown quantum state cannot be
copied. That means that the classical error correction scheme consisting on
preparing several copies of a bit and taking periodic ``majority votes'' to
discard the occasional noisy bit is not usable. Instead, a known relation is
established among a group of redundant qubits which actually only contain one
qubit of useful information. Then, after an error, this relation can be
restored without perturbing the actual value of the qubit. This recovery of the
quantum information without performing a projective measurement must be an
essential feature of any scalable quantum information processing design.

For the implementation of scalable Quantum Information Processing Devices,
molecular electron spin qubits are very promising. Indeed, $g$-tensor
engineering has been achieved to prepare an electron spin-qubit version of a
Lloyd (ABC)$_n$ model.\cite{Lloyd,Morita} Recently, the coherence time of
molecular rare-earth complexes has been extended by means of chemical design
and/or optimal experimental conditions, allowing a high number of coherent Rabi
oscillations.~\cite{Hill,Helena}
Scalability would be achieved by combining these solid state qubits with qubits
of other nature e.g. photons or micro-SQUIDs. This possibility is getting
closer with the advances in heterogeneous quantum information
processing.\cite{MortonARCMP2011} 

It is known that molecules containing more than one spin can be used for
Quantum Error Correction (QEC), i.e. complex molecules can be designed to
function as a single encoded qubit. Except for a phase factor which
does not affect any observable, selective $\pi/2$ and $\pi$ pulses are
available both for the Hadamard and for the CNOT gate in pulsed magnetic
resonance spectroscopy. Indeed, this line of research has been developed for
the last 15 years.\cite{NMRQECsinceCoryPRL1998, Morton2005, SatoJMC2009}

So far, these efforts have been largely limited to nuclear spins, while
magnetic metal complexes have --save exceptions--~\cite{Loss} been used in the
context of Quantum Computing just as proof-of-principle toy
models\cite{LuisPRL2011} or to obtain enhanced properties.\cite{MeierPRL2003}
In the near future, the avenue of electronic spins needs to be explored, since the
coupling of electron spins with qubits of different nature holds great promise.
As recent notable examples, superconducting qubits have been interfaced with
telecom photons via rare-earth complexes~\cite{OBrien2014} and a rare earth has
been locally implanted in a superconducting micro-resonator without degrading
its internal quality factor.~\cite{Wisby2014} In the mid term, there is no
fundamental reason that impedes this avenue leading to molecular (i.e.:
tuneable) analogues of the NV centers in diamond, where QEC has also been
recently demonstrated.~\cite{NVQEC} In this work, we suggest that some
polynuclear metal complexes can also be used to construct non-trivial
building blocks capable of Quantum Error Correction.

\begin{figure}
\begin{tabular}{cc}
\includegraphics[width=0.22\textwidth]{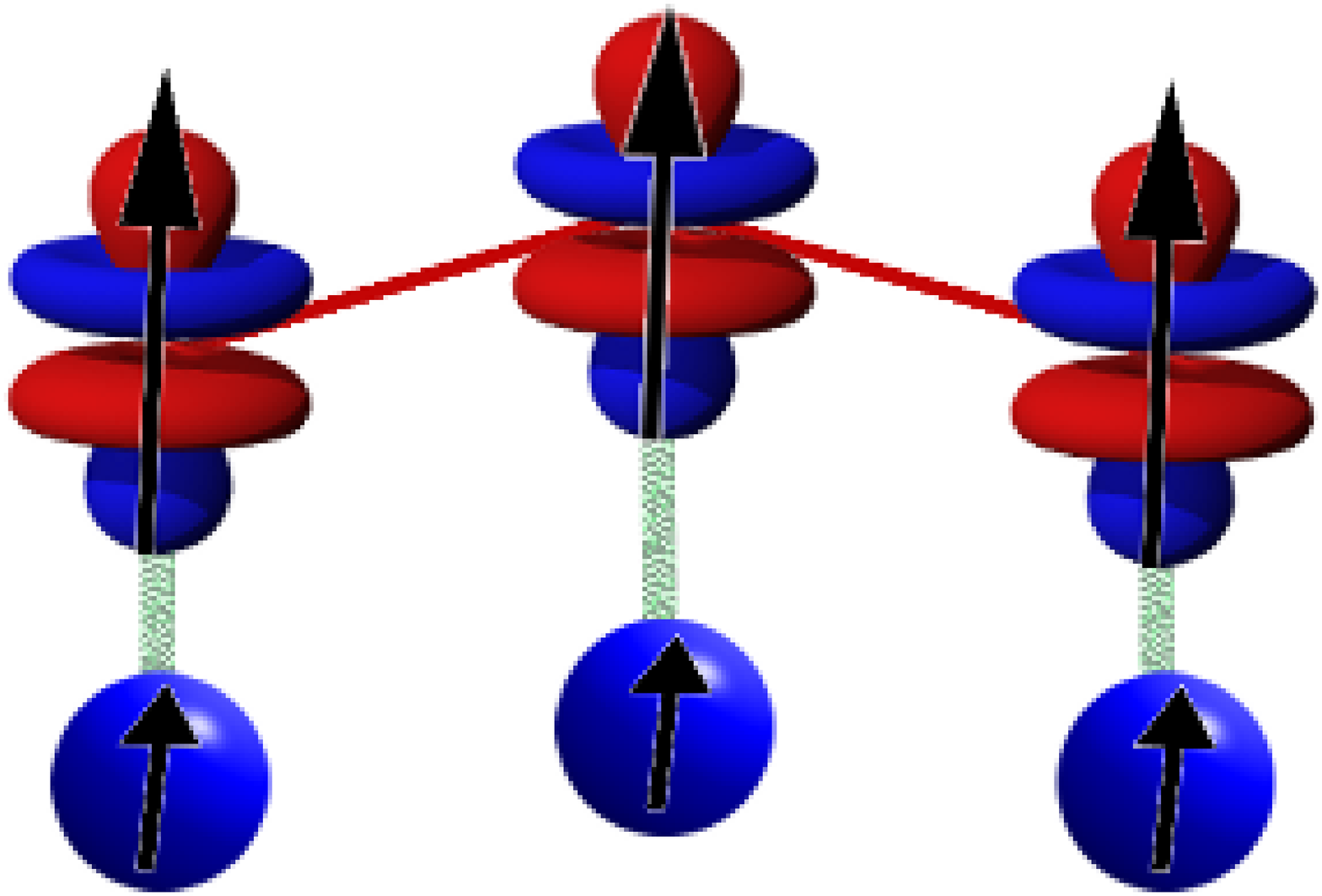}&
\includegraphics[width=0.40\columnwidth]{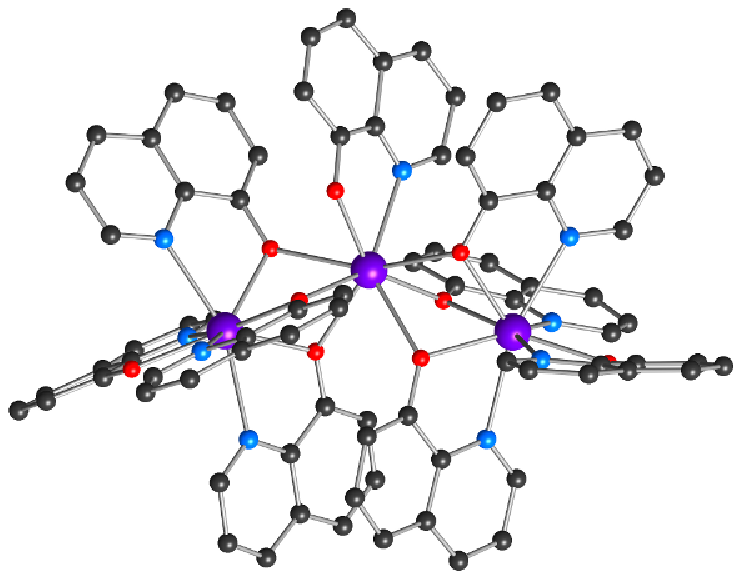} \\
\end{tabular}
\caption{(color online) (left) Interaction scheme between three electronic spins
(doublet ground state) and three nuclear spins quadruplet($I=(3/2)$), resulting
in $2^3\cdot4^3=512$ states or 9 qubits. (right) An experimental example of 
Tb$^{3+}$ trimer, Tb$_3$(OQ)$_9$.}
\label{scheme}
\end{figure}

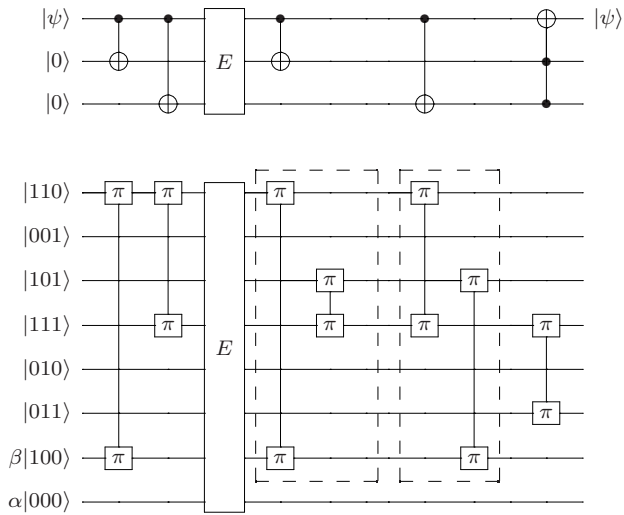
\begin{figure}
\begin{footnotesize}
\begin{align*}
 \Qcircuit @C=1.0em @R=1.0em {
   \lstick{\ket{\psi}}  &\ctrl{1}  &\ctrl{2}  &\multigate{2}{E} &\ctrl{1} & \qw& \qw& \qw &\ctrl{2} & \qw& \qw&\targ     & \rstick{\ket{\psi}} \qw \\
   \lstick{\ket{0}}     &\targ     &\qw       &\ghost{E}        &\targ    & \qw& \qw& \qw &\qw      & \qw& \qw&\ctrl{-1} & \rstick{            } \qw \\
   \lstick{\ket{0}}     &\qw       &\targ     &\ghost{E}        &\qw      & \qw& \qw& \qw &\targ    & \qw& \qw&\ctrl{-2} & \rstick{            } \qw \\
    \\
    \\
   \lstick{\ket{110}}      &\gate{\pi}\qw &\gate{\pi}\qw &\multigate{7}{E}&\gate{\pi}\qw &\qw           &\qw&\qw&\gate{\pi}\qw &\qw           &\qw&\qw           &\qw\\
   \lstick{\ket{001}}      &\qw       \qwx&\qw       \qwx&\ghost{E}       &\qw       \qwx&\qw           &\qw&\qw&\qw       \qwx&\qw           &\qw&\qw           &\qw\\
   \lstick{\ket{101}}      &\qw       \qwx&\qw       \qwx&\ghost{E}       &\qw       \qwx&\gate{\pi}\qw &\qw&\qw&\qw       \qwx&\gate{\pi}\qw &\qw&\qw           &\qw\\
   \lstick{\ket{111}}      &\qw       \qwx&\gate{\pi}\qwx&\ghost{E}       &\qw       \qwx&\gate{\pi}\qwx&\qw&\qw&\gate{\pi}\qwx&\qw       \qwx&\qw&\gate{\pi}\qw &\qw\\
   \lstick{\ket{010}}      &\qw       \qwx&\qw           &\ghost{E}       &\qw       \qwx&\qw           &\qw&\qw&\qw           &\qw       \qwx&\qw&\qw       \qwx&\qw\\
   \lstick{\ket{011}}      &\qw       \qwx&\qw           &\ghost{E}       &\qw       \qwx&\qw           &\qw&\qw&\qw           &\qw       \qwx&\qw&\gate{\pi}\qwx&\qw\\
   \lstick{\beta\ket{100}} &\gate{\pi}\qwx&\qw           &\ghost{E}       &\gate{\pi}\qwx&\qw           &\qw&\qw&\qw           &\gate{\pi}\qwx&\qw&\qw           &\qw\\
   \lstick{\alpha\ket{000}}&\qw           &\qw           &\ghost{E}       &\qw           &\qw           &\qw&\qw&\qw           &\qw           &\qw&\qw           &\qw
   \gategroup{6}{5}{12}{7}{1.0em}{--}
   \gategroup{6}{9}{12}{10}{1.0em}{--}
 }
\end{align*}
\end{footnotesize}
\caption{ (up) Qubit circuit for a 3-qubit QEC scheme correcting a {\it bit
flip} error E, part of the Shor code. (down) Scheme for a pulse sequence between
eigenstates (labels correspond to their logical states) highlighting pairs of
$\pi$-pulses that correspond to a single CNOT gate.}
\label{pulses}
\end{figure}

The most conceptually simple --while general-- implementation of this strategy is
the Shor code,~\cite{Shor} which belongs to the Bacon-Shor error correcting
code class. This class of codes tend to have simpler correction circuits, which
increases the likelihood of finding an experimental system where they can be
carried out. Moreover, Bacon-Shor codes for a given number and kind of errors
can in general be adapted roughly preserving their structure. This opens the
possibility of in-situ adapting the error correction to the nature of the
noise in the physical system.

In this work, we will show that the Shor code could be implemented by using
electron-nuclear Bell states of three magnetic ions. A well-isolated electron
doublet is easily achievable in lanthanoids which would contribute one qubit
per ion;\cite{SIMs} to reach the required nine qubits we
would need two additional qubits per nuclear spin. Therefore,
$^{159}$Tb$^{3+}$, with $I=\frac{3}{2}$ and a 100\% natural isotopic purity,
would be the perfect candidate for this goal, assuming the lowest electronic
doublet has a sufficiently large gap $\Omega$ to the next excited level.
Alternatively, one could use a trimer of $^{63}$Cu$^{2+}$ (or
$^{65}$Cu$^{2+}$), also with $I=\frac{3}{2}$ and where it is also trivial to
achieve a spin doublet (other transition metal candidates exist but are less
convenient).  In any case, the three nuclear quadruplets combined with the
three electronic doublets would provide a $d=2^3\cdot4^3=2^9$ Hilbert space
i.e. the basis of 9 qubits. 

{\it Definition of the system} --
Let us begin by considering the low-energy spectrum of three coupled $^{159}$Tb$^{3+}$
ions, i.e. the full $4^3$ states resulting from the three $I=3/2$ nuclear spins
and the lowest $2^3$ substates of three $J_{L+S}=6${} electronic spins
(Fig.~\ref{scheme}).  Note that effective spins $S=1/2$ which are commonly
used to reproduce spin qubits have two shortcomings in this situation: (1) they
cannot have tunneling splitting, an important feature of non-Kramers rare
earth systems and (2) the magnetic coupling of two $S=1/2$ produces a different
energy level scheme: a triplet plus a singlet instead of the expected
doublet-plus-two-degenerate-singlets.  Thus, we approximate the low-energy
doublet of these $J_{L+S}=6$ electronic states by $S=1$ spins with effectively
infinite axial zero-field splittings --and therefore also infinite gap $\Omega$
to the first excited level-- to produce the correct energy level scheme and
tunneling splitting. 
%
%
%
%

We apply the following Hamiltonian:
\begin{multline}
\hat{H}=-2J_{ex}\left(\hat{S_1}\hat{S_2}+\hat{S_2}\hat{S_3}\right)+\sum_{i=1}^3 \left(D\hat{S_z}_i^2+E(\hat{S_x}_i^2-\hat{S_y}_i^2)\right) \\
+\sum_{i=1}^3 \left( A\hat{S_i}\hat{I_i}+P\hat{I_z}_i^2 +H_z(\mu_B g_e\hat{S_i}_z+\mu_I g_I\hat{I_i}_z)\right)
\end{multline}
where 
$J_{ex}$ is the magnetic exchange, 
$\hat{S}_i$ are effective electron $S=1$ spins,
$\hat{I}_i$ are nuclear $I=3/2$ spins,
$D$ is the linear zero-field splitting, 
$E$ is the rhombic zero-field splitting, 
$A$ is the hyperfine coupling, 
$P$ is the nuclear quadrupole term,
$g_e$ is the effective electronic Land\'e{} factor,
$g_I$ is the effective nuclear Land\'e factor,
$\mu_B$ and $\mu_I$ are the Bohr magneton and the nuclear magneton, 
and $H_z$ is the external field.

For Tb$^{3+}$ we assume common values for hyperfine coupling $A=0.1038$ cm$^{-1}$,
nuclear quadrupole term $P=0.01$ cm$^{-1}$ and effective nuclear Land\'e{}
factor $g_I=0.00073$. We are employing an effective $S=1$ instead of the
typical $S=1/2$, resulting in an effective Land\'e{} factor $g_e=8.915$ which
is equivalent to a typical effective Land\'e{} factor $g_{e(S=1/2)}=17.830$ for
Terbium. 
In our calculations, we explore a certain range of magnetic exchange $J_{ex}$,
rhombic zero-field splitting $E$ (accounting for tunneling splitting $\Delta$)
and external field $H_z$, as these parameters can be varied with a certain
ease, either by chemical or experimental design.  In particular, we explore a
three-dimensional parameter space defined by $-1.0 \leq 2J_{ex} \leq -0.2$
cm$^{-1}$, $0.0 \leq 2E \leq 1.0$ cm$^{-1}$, $0.00 \leq H_z \leq 0.25$ T.
Note that here $J_{ex}$ is acting on an effective $S=1$. The range of both
magnetic exchange and tunneling splitting $\Delta=2E$ corresponds to typical
values among lanthanoid complexes. We simplified the exploration of the
three-dimensional parameter space by choosing the $H_z=f(J_{ex})$ surface such
that the splitting corresponding to a typical Zeeman electronic transition
corresponds to a standard W-band apparatus 95 GHz (other transitions would of
course require a non-standard setup with additional frequencies, see below).
This corresponds to $0.09 \leq H_z \leq 0.22$ T.  We diagonalize the
Hamiltonian with the MAGPACK software package.~\cite{MAGPACK}

In a second independent exploration, we consider three exchange-coupled
$^{63}$Cu$^{2+}$ ($S=1/2)$. This gives us the opportunity to explore a
different parameter space, namely
$0.005<A_\parallel<0.030$~cm$^{-1}$, $A_\perp=0.002$~cm$^{-1}$, $ -10.00$
cm$^{-1} < J_{ex} < -1.00 $ cm$^{-1}$, $P = 0.00127$ cm$^{-1}$, $g_I = 0.00081$,
$g_e^{xy} = 2.0 $, $g_e^z = 2.1$. In this case a typical Zeeman electronic
transition in the range of a W-band spectrometer is achieved by a magnetic
field of $H_z=2.25$ T.  The main difference to the above treatment is that
$S_{Cu^{2+}}=1/2$ therefore $D=E=0$.

{\it General scheme} --
To carry out the algorithm in an ENDOR setup, the logical quantum circuit needs
to be translated into a pulse sequence (Fig.~\ref{pulses}).  Additionally, we
need to consider that Quantum Error Correction using the Shor code only makes
sense if all allowed transitions at the error step correspond to single-qubit
errors. Finally, a mechanism for readout as a final step of the algorithm must
be in place. As described below, all three requirements can be met by a proper
assignment of spin-qubit labels.

For a practical assignment of qubit labels to spin states, it is ideal to
separate the six nuclear spin-qubits from the three electronic spin-qubits. The
512 electronuclear states defining our 9 qubits are, however, not necessarily
separable, owing to the hyperfine interaction which results in a certain degree
of mixing. As a numerical test for this separability, we calculate the fidelity
$F_{as}= |\langle\Psi_a|\Psi_s\rangle|$ between the (actual) states $\Psi_a$ with
a given set of parameters and the (simplified) states $\Psi_s$ that would
result from the cancellation of the hyperfine coupling. As the energy order
changes, for each $\Psi_a$ we choose $\Psi_s$ that maximizes $F_{as}$.
Note that in the case of Tb, the effectively infinite (negative) value of $D$
means that, while we produce $3^3\cdot4^3=1728$ states, we work only with the
lowest 512, the other 1206 being beyond our ultraviolet cutoff.  

\begin{figure}[ht]
\includegraphics[width=0.49\columnwidth]{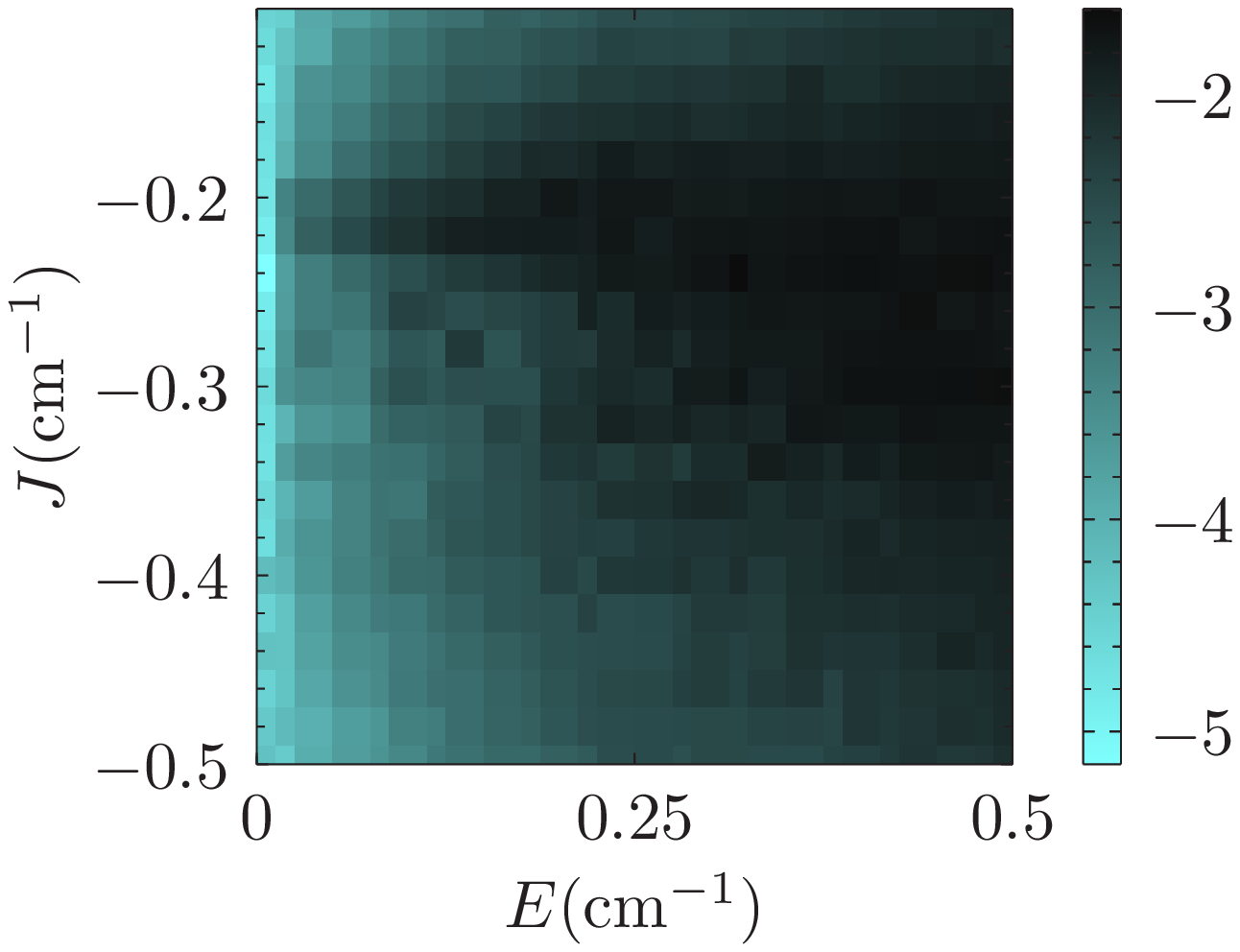}
\includegraphics[width=0.49\columnwidth]{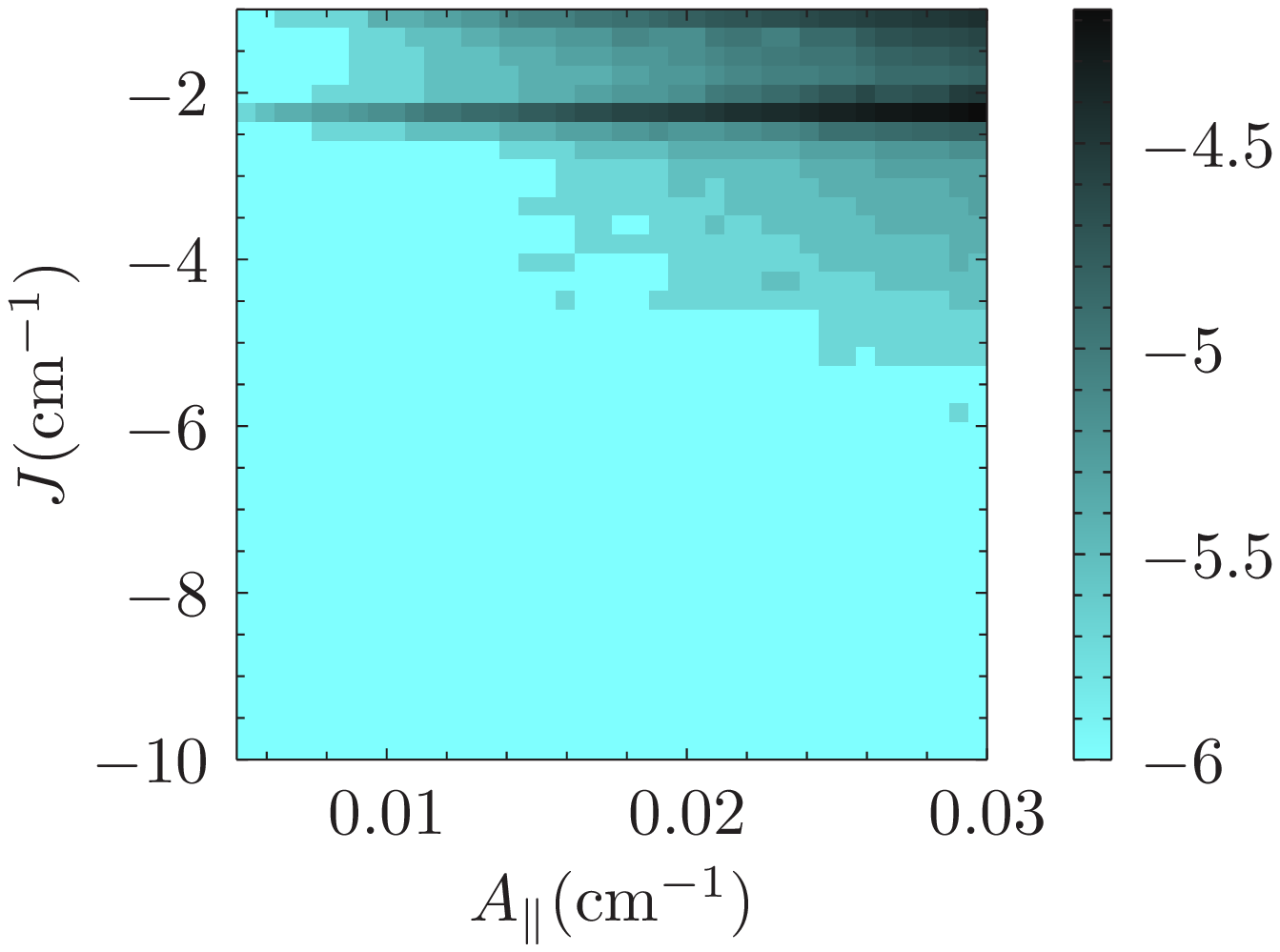}
\caption{(color online) Variation of log$(1-F_{as})$ where $F_{as}$ is the
128$^{\rm th}$ highest fidelity as a function of the tunneling splitting
$\Delta$, the magnetic exchange $J_{ex}$ and the external field $H_z$. Left:
Tb$^{3+}$ system; right: Cu$^{2+}$ system (see text). $F_{as}=10^{-6}$ is the
precision limit of our calculation.}
\label{fidelity}
\end{figure}

Note that not all 512 lowest states are actually needed. The encoded state employs
superpositions of all 8 electronic states but only 8 (of 64) nuclear states,
resulting in 8 (of 512) electronuclear states. In fact, until the error
occurs the full evolution of the system can be described with a mere 22
electronuclear states. Furthermore, with the safe assumption that the error
will occur in the electronic part of the system --characterized by faster
transitions and shorter decoherence times-- the evolution of the system in most
experiments will be limited to less than 100 different states. Thus, less than
100 (of 512) different qubit labels are employed from the start to the end of
the algorithm. As a conservative estimate, we plotted the bottom of the first
quartile (128$^{\rm th}$ highest fidelity).  As seen in Fig.~\ref{fidelity}, in
both cases the obtained fidelities are very high in ample parameter areas:
$(1-F_{as})<10^{-4}$ for Tb$^{3+}$, $(1-F_{as})<10^{-6}$ for Cu$^{2+}$.
This demonstrates the feasibility of the separation of the electronic and
nuclear parts of the wave function, and thus the independent assignment of
spin-qubit labels.  

The next task is guarantee that in the encoded state all transitions that are
allowed both by spin and by symmetry --the most likely errors-- correspond
to single-qubit errors, thus correctable by the Shor code. This is done by a
proper assignment of qubit labels to spin states.
Table~\ref{electronbasis} indicates the correspondence of the electronic
eigenstates and their qubit values for three effective $S=1/2$ (or the lowest
part of the spectrum in the case of an $S=1$ as discussed above) with a linear
connectivity. It was built to guarantee that (a) all spin- and symmetry-allowed
transitions from the encoded state correspond to single-qubit errors and can
thus be corrected and (b) the value of the first qubit can be inequivocally read
from a measurement determining the absolute value of $M_s$.  Assuming errors
will occur in the electronic spin, for the nuclear spins there are many valid
spin-qubit mappings.

\begin{table}[hb]
\caption{Correspondence between electronic qubits, quantum numbers and wave
function. $A>a$, exact values depending on $J_{ex}$.}
\begin{tabular}{c|c|c|c}
$\ket{ABA}$ & $S$ & $M_s$ & $\Psi$             \\
\hline      
$\ket{111}$ & 1/2 & -1/2  &  $A\ket{\downarrow\uparrow\downarrow}-a\left(\ket{\uparrow\downarrow\downarrow}+\ket{\downarrow\downarrow\uparrow}\right)$ \\
$\ket{011}$ & 1/2 & +1/2  &  $A\ket{\uparrow\downarrow\uparrow}-a\left(\ket{\downarrow\uparrow\uparrow}+\ket{\uparrow\uparrow\downarrow}\right)$ \\
$\ket{101}$ & 1/2 & -1/2  &  $\ket{\downarrow\downarrow\uparrow}-\ket{\uparrow\downarrow\downarrow}$ \\
$\ket{001}$ & 1/2 & +1/2  &  $\ket{\downarrow\uparrow\uparrow}-\ket{\downarrow\uparrow\uparrow}$ \\
$\ket{100}$ & 3/2 & -3/2  &  $\ket{\downarrow\downarrow\downarrow}$ \\
$\ket{110}$ & 3/2 & -1/2  &  $\ket{\downarrow\uparrow\downarrow}+\ket{\downarrow\downarrow\uparrow}+\ket{\uparrow\downarrow\downarrow}$ \\ 
$\ket{010}$ & 3/2 & +1/2  &  $\ket{\uparrow\downarrow\uparrow}+\ket{\uparrow\uparrow\downarrow}+\ket{\downarrow\uparrow\uparrow}$ \\
$\ket{000}$ & 3/2 & +3/2  &  $\ket{\uparrow\uparrow\uparrow}$ \\
\end{tabular}
\label{electronbasis}
\end{table}

{\it Experimental considerations} --
In the last years, the chemical design of single-molecule magnets and molecular
spin qubits, including clusters of different nuclearity, has been the hottest
topic in molecular magnetism.~\cite{SIMs,SIMdesign} The exchange interaction in
such systems, whether transition metals (superexchange) or lanthanoids (mostly
dipolar) can be tuned in strength and sign. Regarding the chemical design of
molecules that would be appropiate for this scheme, the versatility of
coordination chemistry allows their preparation connected either through
covalent bonds (i.e. formation of a discrete molecule) or via supramolecular
interactions, such as hydrogen bonds. Figure~\ref{scheme}(b) shows a recent
example of the covalent class of trimers, but in general the exchange
interaction can be controllably weakened from the covalent situation to the
supramolecular case e.g. [\{Tb(TETA)\}$_2$Tb(H$_2$O)$_8$]$^+$. The case of
Cu$^{2+}$ is much more favourable from the point of view of the electron-nuclear
separability but presents some complications, in particular the difficulty to
magnetically isolate the trimers from each other. We will center this
discussion in the Tb$^{3+}$ trimer.


As a model lanthanoid system, we study {\bf 1} with the software package
SIMPRE\cite{SIMPRE} using an effective point charge model to obtain the ground
state wave function, the tunneling splitting $\Delta$, the gap to the first
excited state $\Omega$ and $g_x, g_y, g_z$ (Table~\ref{tablesimpre}). The
results justify our assumptions in the order of magnitude of the relevant
parameters.

\begin{table}[h]
\caption{Single-ion characterization of the Tb$^{+3}$ ions in {\bf 1}:
composition of the ground-state wave function, tunneling splitting, gap to the
first excited state, anisotropic Land\'e $g$-factor. Note that Tb1 and Tb3 are
similar but not crystallographically equivalent.}
\begin{tabular}{c|c|c|c|c|c|c}
ion & $\Psi_{GS}$  & $\Delta({\rm cm}^{-1})$ & $\Omega({\rm cm}^{-1})$ & $g_z$ & $g_x$ & $g_y$ \\
\hline
Tb1 & $96\%|\pm6>$ &        0.86             &      89.7               &17.109 & 0.951 & 1.095 \\
Tb2 & $96\%|\pm6>$ &        0.92             &      191.2              &17.196 & 0.539 & 0.717 \\
Tb3 & $96\%|\pm6>$ &        1.28             &      83.7               &16.807 & 0.951 & 1.132 \\
\end{tabular}
\label{tablesimpre}
\end{table}

In a target system, the initialization to zero i.e. the preparation of the
ground state $\ket{000000000}$ can be achieved by cooling at mK temperatures.
Alternatively, one can work with pseudo-pure states,~\cite{pseudopure} as
is routinely done in NMR quantum computing setups.  The `writing' of the
non-trivial qubit starting state would then be an arbitrary, coherent
transition between the states $ |\Psi_0\rangle =
|000000000\rangle $ and $ |\Psi_1\rangle =
|100000000\rangle
$ i.e. the electron part of the wave function is rotated via three consecutive
microwave pulses between $\ket{\uparrow\uparrow\uparrow}$ and
$\ket{\downarrow\downarrow\downarrow}$
by desired amount, preserving the nuclear part.  Note that the negation of the
``target'' qubit can be seen as a full transition or $\pi$ EPR pulse
and Hadamard gate is simply a $\pi/2$ pulse that -for a single qubit-
transforms $\ket{0}$ into $(1/\sqrt{2})(\ket{0}+\ket{1}))$ and $\ket{1}$ into
$(1/\sqrt{2})(\ket{0}-\ket{1}))$, i.e. transforms `bit' information into `phase'
information.  Thus, a series of allowed transitions can encode this state
either using just the three electronic qubits for the correction of a single
error type (Fig. 1), or using the nine electronuclear qubits for Shor's QEC
code.  

A complete measurement of the final state of the system can be made simply
through a Electron Spin Echo (ESE) i.e. the detection of the standard pulsed
EPR signal. Different final states have different ESE spectra, which can be
compared with simple states prepared in an independent experiment. If the error
is introduced as a controlled operation, all copies of the molecule in the
ensemble will share the same state. If the error happened by itself, the ESE
will result of the superposition of the different possibilities, weighted
according to each error rate. Interestingly, in the chosen spin-qubit
labelling scheme, it is possible to experimentally measure the value
of the first qubit just by determining the sign of $M_s$ (i.e. `is the sample
attracted or repelled by a given external magnetic field?'). 

The full procedure of the Shor code is conceptually simple to derive but
experimentally will require the ability to apply pulses of many different
frequencies, which can be achieved by an Arbitrary Waveform Generator.
Note that the preparation and manipulation of the pseudo-pure state 
could be implemented using the time-proportional-phase-increment technique,
which combines pulses on electron and nuclear spins with waiting times to
cancel the off-diagonal term of the density matrix. Special rotation
angles need to be applied to the detection pulses in order to distinguish
entangled states from superposition states.~\cite{SatoJMC2009} Realisation of
entanglement between the qubits encoded in an electron spin and a nuclear spin
has already been reported in an ENDOR experiment using pseudo-pure
states.~\cite{Mehring03,Mehring04}

Of course, a minimalistic alternative to the full Shor code would be just to
use the electrons spins as three qubits for either spin flip or phase flip
errors. This could be done with no interference from nuclear states using
$S=1/2$, $I=0$ complexes, which can be achieved with either lanthanoids
($^{164}$Dy, $^{166}$Er) or transition metals (low-spin $^{56}$Fe$^{3+}$).
More sophisticated alternatives include using electron spins as ``bus spins''
and nuclear spins as ``client qubits'', i.e. encoding a quantum state on the
electron spin, then transferring it for protection to the nucleus until
retrieval is necessary. This approach, which has not yet been implemented
in molecular systems, profits from the longer decoherence time of nuclear spins
and the faster operation capability of electron spins.

{\it Concluding remarks} --
In this contribution we explore the possibility of using certain magnetic
molecules as dense clusters of electronuclear qubits for Quantum Error
Correction. We show that the magnetic coupling between electron spins forces a
non-trivial spin-qubit label mapping. Note that always-on coupling is common in
the solid state e.g. NV centers also have always-on hyperfine coupling, thus
having tools to deal with it opens new materials as candidate hardware. Exploring a
realistic parameter space, we demonstrate that the nuclear part of the spin
wave function can be practically decoupled from the electronic part by chemical
design, simplifying the implementation of the pulsed EPR experiment. An
important advantage of our approach is that it favours non-correlated noise, as
we are free to assign the spin-qubit labeling in a way which ensures that
transitions flipping more than one qubit are forbidden transitions, and thus
statistically unlikely. Hopefully, this work will stimulate further advances in
the field of heterogeneous quantum information processing.

\section{Acknowledgements}
We thank Joris van Slageren for a crucial comment. The present work has been
funded by the EU (Project ELFOS and ERC Advanced Grant SPINMOL), the Spanish
MINECO (grant MAT2011-22785 and the CONSOLIDER project on Molecular
Nanoscience), and the Generalitat Valenciana (Prometeo and ISIC Programmes of
excellence). A.G.A. acknowledges funding by the MINECO (Ram\'on y Cajal
Program), J.J.B thanks the MINECO for an FPU predoctoral grant.

\section{Author contributions}
JJB and LEM did the single-ion work. LEM and SCS did the 3- and 9-qubit work.
LEM, JMCJ and GME contributed crucial ideas. JMCJ and LEM wrote the software.
AGA conceived and supervised the project and wrote the paper. All authors
reviewed and contributed to the manuscript.

\end{document}